\documentclass[letterpaper, 10pt, journal, twoside]{IEEEtran}
\usepackage{cite}
\usepackage{amsmath,amssymb,amsfonts}
\usepackage{algorithm}
\usepackage{algpseudocode}
\usepackage{graphicx}
\usepackage{textcomp}
\usepackage{hyperref}
\usepackage{makecell}

\IEEEoverridecommandlockouts

\usepackage{amsthm}

\usepackage{mathtools}

\newtheorem{assumption}{\bf Assumption}

\newtheorem{proposition}{\bf Proposition}
\newtheorem{theorem}{\bf Theorem}

\newtheorem{remark}{Remark}
\usepackage{xcolor}

\newcommand{\R}{{\mathbb{R}}}
\newcommand{\E}{{\mathbb{E}}}

\newcommand{\I}[1]{\mathbb{I}_{[#1]}}

\DeclareMathOperator*{\argmin}{arg\,min}

\def\changes{\textcolor{black}} 

\makeatletter
\newcommand\fs@betterruled{%
  \def\@fs@cfont{\bfseries}\let\@fs@capt\floatc@ruled
  \def\@fs@pre{\vspace*{5pt}\hrule height.8pt depth0pt \kern2pt}%
  \def\@fs@post{\kern2pt\hrule\relax}%
  \def\@fs@mid{\kern2pt\hrule\kern2pt}%
  \let\@fs@iftopcapt\iftrue}
\floatstyle{betterruled}
\restylefloat{algorithm}
\makeatother

\begin{document}
\title{\LARGE \bf Stochastic Data-Driven Predictive Control:\\Chance-Constraint Satisfaction with Identified Multi-step Predictors}

\author{Haldun Balim$^{1,2}$, Andrea Carron$^1$, Melanie N. Zeilinger$^1$, Johannes Köhler$^1$ \thanks{$^1$Institute for Dynamic Systems and Control, ETH Zurich, Switzerland.}
\thanks{$^2$School of Engineering and Applied Sciences, Harvard, Massachusetts, USA (e-mail: hbalim@fas.harvard.edu).}
\thanks{This work has been supported by the Swiss National Science Foundation
under NCCR Automation (grant agreement 51NF40 180545) and the ETH Career Seed Award funded through the ETH Zurich Foundation.}
}

\maketitle
\thispagestyle{empty}
\pagestyle{empty} 

\begin{abstract}
We propose a novel data-driven stochastic model predictive control framework for uncertain linear systems with noisy output measurements. Our approach leverages multi-step predictors to efficiently propagate uncertainty, ensuring chance constraint satisfaction. In particular, we present a strategy to identify multi-step predictors and quantify the associated uncertainty using a surrogate (data-driven) state space model. Then, we utilize the derived distribution to formulate a constraint tightening that ensures chance constraint satisfaction \changes{despite the parametric uncertainty}. A numerical example highlights the reduced conservatism of handling parametric uncertainty in the proposed method compared to state-of-the-art solutions.

\end{abstract}
\begin{IEEEkeywords}
Predictive control for linear systems, Data driven control, Identification for control
\end{IEEEkeywords}

\section{Introduction} \label{sec:intro}
Ensuring satisfaction of safety-critical constraints while controlling uncertain systems presents a significant challenge~\cite{hou2013model}. Model Predictive Control (MPC) is an effective methodology to tackle this issue, due to its intrinsic capabilities to handle constraints and address complex multi-input multi-output systems~\cite{kouvaritakis2016model}. Implementing an MPC scheme requires a prediction model, yet acquiring an accurate model is difficult~\cite{ogunnaike1996contemporary}. 
This has sparked considerable interest, focusing on both direct~\cite{berberich2020data, coulson2021distributionally, breschi2023data} and indirect~\cite{terzi2019learning, koehler2022state} data-driven predictive control approaches, i.e., control methods that primarily utilize data for predictive controller design. However, enforcing safety-critical constraints when the data is subject to noise remains as a significant challenge. The proposed framework addresses this problem by identifying \textit{multi-step predictors} to directly enforce probabilistic constraints.


\subsection{Related Work} 
While data-driven methods provide significant advantages, they also introduce unique challenges, especially in predictive control applications~\cite{hou2013model}, where quantifying uncertainty in predictions complicates the design process.

\textit{State-space models:} 
One of the most natural approaches to address this problem is the identification of a state-space model and the subsequent design of an MPC formulation that robustly accounts for parametric uncertainty, see, e.g., ~\cite{balim2024data, arcari2023stochastic}.
These methods typically ensure satisfaction of safety-critical constraints by using an efficient sequential propagation of the uncertainty.
However, this sequential propagation often leads to overly conservative solutions. 

\textit{Multi-step predictors:} Recently, various approaches have been proposed that leverage direct multi-step ahead prediction to circumvent issues arising from the sequential propagation of parametric uncertainty in state-space models~\cite{berberich2020data,terzi2019learning, koehler2022state,coulson2021distributionally}.

Particularly, \textit{direct} data-driven approaches have recently gained significant traction~\cite{coulson2021distributionally, verheijen2023handbook}. These methods utilize input-output data to implicitly construct multi-step predictors, bypassing traditional model identification. However, extending this strategy to stochastic systems often hinges on strong assumptions, e.g., that a disturbance sequence associated with offline data is precisely known~\cite{yin2023stochastic, coulson2021distributionally, wang2022data,pan2022stochastic, huang2023robust}. 

Alternatively, \textit{indirect} data-driven methods first identify  multi-step predictors from data~\cite{koehler2022state,terzi2019learning,saccani2023homothetic,furieri2022near}. Notably, the strategy of parameterizing multi-step transitions as distinct models has already been explored in seminal works~\cite{zheng1993robust} in the form of finite impulse response models. Nevertheless, the identification of general multi-step predictor models and their subsequent use for robust/stochastic predictive control purposes pose substantial challenges. 
In case of bounded noise and disturbances, set-membership identification can be leveraged to characterize multi-step predictors from data~\cite{terzi2019learning}. 
Given such bounds on multi-step predictors, robust control designs can be directly leveraged to ensure satisfaction of constraint for worst-case uncertainty~\cite{saccani2023homothetic, terzi2019learning,furieri2022near}.
However, such \textit{robust} methods can lead to overly conservative results and in general are not suitable for rare outlier noise. 
In contrast,~\cite{koehler2022state} proposes a strategy to identify multi-step predictors in case of unbounded \textit{stochastic} disturbances with noisy state measurements. Additionally, a constraint tightening approach is introduced to ensure satisfaction of chance constraints, employing ellipsoidal confidence bounds for the model parameters. Finally,~\cite{fiedler2023} presents a strategy for identifying multi-step predictors for stochastic systems and introduces a constraint tightening method. However, the approach neglects parametric uncertainty in the propagation of state covariance. Overall, ensuring chance-constraints in data-driven predictive control with stochastic output measurements remains an open problem.

\subsection{Contribution and Outline} 
The primary contribution of this paper is the development of a data-driven predictive control framework that employs multi-step predictors to ensure chance-constraint satisfaction for linear stochastic systems using only input-output data. First, we define the considered stochastic optimal control problem in Section~\ref{sec:setup}. In Section~\ref{sec:solution}, we outline the proposed data-driven solution. Specifically, we provide a novel method for identifying multi-step predictors and quantifying the associated uncertainty, which will be discussed in Section~\ref{sec:id}. This method applies \changes{Maximum Likelihood Estimation (MLE)} based on the \textit{innovation form} of the system obtained through Kalman filter recursions of an estimated state-space model. Afterwards in Section~\ref{sec:control}, the proposed predictive control problem ensures chance constraint satisfaction by accounting for the uncertainty of the estimated parameters. 
Overall, the proposed solution offers an alternative strategy to propagate the epistemic uncertainty of data-driven state-space models. 
Finally, in Section~\ref{sec:example}, we validate our approach with a numerical example, demonstrating reduced conservatism compared to state-of-the-art solutions, and conclude the paper in Section~\ref{sec:conclusion}.

\subsection{Notation} 
The set of integers within the interval $[a,\ b] \subseteq \R$ is represented by $\I{a,b}$. For a sequence of matrices $X_t \in \R^{n\times m}$, with $t \in \I{t_1,t_2}$ and $t_1,t_2 \in \I{\geq 0}$, the stacked matrix is denoted by $X_{[t_1:t_2 ]} \in \R^{n(t_2 + 1 - t_1)\times m}$. We use $I$ to denote identity matrix of appropriate dimensions. For a square matrix $A \in \R^{n \times n}$, we use $A^{1/2}$ to denote the symmetric square-root. The expected value of a random variable $x$ and the probability of an event $A$ are denoted by $\E[x]$ and $\Pr[A]$, respectively. For conditional probability of $A$ given event $B$ we use $\Pr[A\mid B]$. A variable $x$ following a Gaussian distribution with mean $\mu$ and variance $\Sigma$ is denoted as $x \sim \mathcal{N}(\mu, \Sigma)$. The quantile function of the chi-squared distribution with $n$ degrees of freedom at probability $p \in [0, 1]$ is denoted by $\chi^2_n(p)$. The positive semi-definiteness of a matrix $Q$ is denoted by $Q \succeq 0$, and we write $\|x\|_Q^2 = x^\top Qx$ for $Q \succeq 0$. The Kronecker product is represented by $\otimes$ and $\mathrm{vec}(A) \in \mathbb{R}^{nm}$ denote the operation that transforms a matrix $A \in \mathbb{R}^{n \times m}$ into a vector by sequentially stacking its columns. To denote a block-diagonal matrix with matrices $A$ and $B$ on the diagonal, we use $\mathrm{diag}(A, B)$. Similarly, to express a block-diagonal matrix with matrix $A$ repeated $n$ times on its diagonal, we use $\mathrm{diag}_n(A)$. \changes{$A \propto B$ indicates direct proportion between terms $A$ and $B$.}

\section{Problem Setup}\label{sec:setup}
We examine discrete-time linear time-invariant systems, which are described as:
\begin{align}\label{eq:state-space}
    x_{t+1} &= A x_t + B u_t + Ew_t, \\
    y_t &= C x_t + v_t,\notag \\
    w_t &\sim \mathcal{N}(0, I),\ v_t\sim\mathcal{N}(0, R) \notag,
\end{align}
with state $ x_t \in \mathbb{R}^{n_{\mathrm{x}}} $, control input $ u_t \in \mathbb{R}^{n_{\mathrm{u}}} $, measured output $ y_t \in \mathbb{R}^{n_{\mathrm{y}}} $, time $ t \in \I{\geq 0} $, disturbance $ w_t \in \mathbb{R}^{n_{\mathrm{w}}} $, and measurement noise $ v_t \in \mathbb{R}^{n_{\mathrm{y}}} $. Both $ w_t $, $ v_t $ are independently and identically (i.i.d.) Gaussian distributed with zero mean and covariance matrices $I$, $R \succ 0$, correspondingly. We consider the following stochastic optimal control problem:
\begin{subequations}\label{eq:soc}
\begin{align}
    &\min_{u_{[0, N-1]} \in \mathbb{U}^N}\: \E \left[\sum_{k=0}^{N-1}\|y_{k+1}\|_{Q_\mathrm{c}}^2 + \|u_k\|_{R_\mathrm{c}}^2\right] \\
    \text{s.t.}  \; 
    &\Pr[h_{j}^\top y_k \leq 1] \geq p,\ j\in \I{1,r},\ k\in\I{1, N},\label{eq:chance-constr}\\
     &x_0 \sim \mathcal{N}(\bar x_0, \Sigma_{\mathrm{x},0}),\ \eqref{eq:state-space} \label{eq:initial-state}.
\end{align}
\end{subequations}
with polytopic input constraint set $\mathbb{U}$, cost weights $Q_\mathrm{c},\ R_\mathrm{c} \succeq 0$, initial state distribution parameters $\bar x_0, \Sigma_{\mathrm{x},0}$, and individual half-space chance constraints~\eqref{eq:chance-constr} with probability $p \in (0, 1)$. When the system matrices $A$, $B$, $C$, $E$ and the noise covariance $R$ are known, Problem~\eqref{eq:soc} can be equivalently written as a quadratic program~\cite{kouvaritakis2016model}. In this paper, we address Problem~\eqref{eq:soc} when $A$, $B$, $C$, $E$, $R$ are largely unknown and need to be identified from data. For the controller design, we have access to an input-output trajectory of length $T\in\I{\geq 1}$ denoted by $ Y_T \coloneqq \{y_t\}_{t=1}^T $ and $ U_T \coloneqq \{u_t\}_{t=0}^{T-1} $ obtained from an open-loop control experiment. For simplicity of exposition, we only optimize over open-loop input sequences in this work.

\section{Solution Approach}\label{sec:solution}
In this section, we outline our strategy to solve the data-driven stochastic optimal control problem~\eqref{eq:soc} using multi-step predictors.

\subsection{Multi-step predictors}
First, we reformulate Problem~\eqref{eq:soc} using multi-step predictors. For any $k \in\mathbb{I}_{[1, N]}$, the $k$-step ahead output-prediction is given by:
\begin{equation}\label{eq:multi-step}
    y_{k} = G_{0, k} x_0 + G_{\mathrm{u},k} u_{[0, k-1]} +  G_{\mathrm{w},k} w_{[0, k-1]} + v_{k}.
\end{equation}
The matrices $G_{\mathrm{0},k}$, $G_{\mathrm{u},k}$, $G_{\mathrm{w},k}$ can be computed based on the system matrices in~\eqref{eq:state-space} as follows:
\begin{align}
    G_{\mathrm{u},k} &= C[A^{k-1}B, \ldots, B],\ G_{\mathrm{0},k} = CA^k, \\ G_{\mathrm{w},k} &= C[A^{k-1}E, \ldots, E]. \notag
\end{align}
Using the parameterization~\eqref{eq:multi-step} with initial condition~\eqref{eq:initial-state}, the output prediction satisfies $y_k\sim \mathcal{N}(\bar y_k, \Sigma_{\mathrm{y}, k})$ with
\begin{align}
     \bar y_k &= G_{\mathrm{0},k} \bar x_0 +  G_{\mathrm{u},k} u_{[0, k-1]},\\
    \Sigma_{\mathrm{y}, k} &= G_{\mathrm{0},k}\Sigma_{\mathrm{x},0} G_{\mathrm{0},k}^\top + G_{\mathrm{w},k} G_{\mathrm{w},k}^\top + R \notag.
\end{align}
 Thus, we can equivalently write Problem~\eqref{eq:soc} as the following convex quadratic program:
\begin{subequations}\label{eq:soc-msp}
\begin{align}
     &\min_{u_{[0, N-1]} \in \mathbb{U}^N} \:  \sum_{k=0}^{N-1}\| \bar y_{k+1}\|_{Q_\mathrm{c}}^2 + \|u_k\|_{R_\mathrm{c}}^2 \\
    \text{s.t.}  \; 
    &h_{j}^\top \bar y_k \leq 1-c_\mathrm{p}\|h_{j}\|_{\Sigma_{\mathrm{y},k}},\ j\in \I{1,r},\\
    &\Sigma_{\mathrm{y}, k} = G_{\mathrm{0},k}\Sigma_{\mathrm{x},0} G_{\mathrm{0},k}^\top + G_{\mathrm{w},k} G_{\mathrm{w},k}^\top + R, \label{eq:covar-dyn}\\
    &\bar y_k= G_{\mathrm{0},k} \bar x_0 +  G_{\mathrm{u},k} u_{[0,k-1]},\ k\in\I{1, N}, \label{eq:baryk}
\end{align}
\end{subequations}
with $c_p = \sqrt{\chi_1^2(2p-1)}$. Note that the optimization problem~\eqref{eq:soc-msp} is an exact reformulation of the stochastic optimal control problem~\eqref{eq:soc}, provided the multi-step predictors $G_{\mathrm{0},k}$, $G_{\mathrm{u},k}$, $G_{\mathrm{w},k}$ are known.

\subsection{Proposed Solution}
Next, we explain how we solve Problem~\eqref{eq:soc-msp} using data-driven estimates of the multi-step predictors. 

\begin{algorithm}
\caption{\changes{General approach}}
\label{algo:solution}
\begin{algorithmic}[1]
\Statex \textbf{Input:} \changes{Input-output trajectory}
 \Statex \% Offline Design:
 \State Estimate state-space model, e.g., using Expectation-Maximization or Prediction Error Method~\cite{GIBSON20051667, Ljung1998}.
 \State \changes{Estimate states and innovation covariances using Kalman filter with estimated state-space model.}
 \State \changes{Estimate multi-step predictors and quantify uncertainty using MLE.}
 \Statex \% Online Phase:
 \State \changes{Solve deterministic optimization problem and apply inputs $u_{[0, N]}$.}
\end{algorithmic}
\end{algorithm}

Notably, direct identification of the long-horizon relationships in multi-step predictors is more challenging than traditional state-space models due to correlations and larger parameter spaces. 

To address this, we propose an identification and uncertainty quantification scheme that leverages the Kalman filter applied with estimated state-space matrices as an intermediate step (Sec.~\ref{sec:id}). With these data-driven multi-step predictors and their associated uncertainties, we develop a predictive control strategy to ensures satisfaction of chance constraints in a \changes{despite the parametric uncertainty}(Sec.~\ref{sec:control}), avoiding the limitations of sequential propagation in state-space models. 
The overall approach is summarized in Algorithm~\ref{algo:solution}.

\section{Parameter Identification for Multi-step Predictors}\label{sec:id}
In this section, we introduce a method to estimate model parameters for the multi-step predictors defined in~\eqref{eq:multi-step}. 

System~\eqref{eq:state-space} can be equivalently written in \textit{innovation form} using Kalman filter recursions as follows~\cite{breschi2023data}:
\begin{subequations} \label{eq:inno-form}
\begin{align}
    x_{t+1\mid t+1} &=  A x_{t\mid t} + Bu_t + L e_t, \label{eq:inno-form-p}\\
    y_{t+1} &= CA x_{t\mid t} + C B u_t + e_{t+1}, \label{eq:inno-form-m}
\end{align}
\end{subequations}
with innovation $e_t \sim \mathcal{N}(0, S)$ i.i.d., (stationary) Kalman gain $L$, innovation covariance $S$, and posterior mean $x_{t\mid t} = \E[x_t \mid y_{[1:t]}]$, all of which can be computed with Kalman filter recursions. For simplicity of exposition, we consider steady-state Kalman gain innovation covariance. The following proposition provides the maximum likelihood estimate for the multi-step predictors, assuming that the Kalman filter ~\eqref{eq:inno-form} is available.
\begin{proposition}\label{prop:id}
    Consider the unknown parameters $\theta_k = \mathrm{vec}([G_{\mathrm{0}, k},\ G_{\mathrm{u}, k}]) \in \R^{n_\theta}$ with $n_{\theta,k}=n_\mathrm{y}(kn_\mathrm{u}+n_\mathrm{x})$ for some $k\in \I{1,N}$, the regressor $\Phi_{k,j} \coloneqq [x_{j\mid j}^\top,\ u_{[j,j+k-1]}^\top]\otimes I_{n_\mathrm{y}} \in \R^{n_\mathrm{y}\times n_{\theta,k}}$, and the covariance matrix $\Sigma_{\mathrm{e}, k}$ according to~\eqref{eq:cov-sigma-e}. Suppose that $\Sigma_{\theta, k}^{-1}\coloneqq \Phi^{\top}_{k,[0,T-k]} \Sigma_{\mathrm{e},k}^{-1} \Phi_{k, [0,T-k]} \succ 0$. Then, the generalized least-squares estimate:
    \begin{align}\label{eq:hat-thetak}
        \hat{\theta}_k \coloneqq & \argmin_{\theta_k} \|y_{[k,T]} - \Phi_{k,[0,T-k]}\theta_k\|^2_{\Sigma_{\mathrm{e},k}^{-1}}
    \end{align}
    is the maximum likelihood estimate for the $\theta_k$ given the output $y_{[k,T]}$ and regressor $\Phi_{k,[0,T-k]}$. 
\end{proposition}
\begin{proof}
First, we apply Eq.~\eqref{eq:inno-form-p} iteratively $k-1$ times starting from time $t$, then predict the output $y_{t+k}$ using Eq.~\eqref{eq:inno-form-m}, resulting in:
\begin{align}\label{eq:msp-trans}
    y_{t+k} &= G_{0, k}x_{t\mid t} + G_{u,k}u_{[t,t+k]} +  G_{\mathrm{e},k} e_{[t,t+k]},
\end{align}
with $G_{\mathrm{e},k} = C[A^{k-1}L,\ \cdots,\ AL,\ I]$. Using properties of Kronecker product Eq.~\eqref{eq:msp-trans} is equivalent to:
\begin{align}\label{eq:e_defn}
    y_{t+k}&= \Phi_{k, t} \theta_k + \tilde e_{t, k},
\end{align}
where $\tilde e_{t, k} = G_{\mathrm{e},k} e_{[t:t+k]}$. Resultantly, $\tilde{e}_{[0:T-k],k} \sim \mathcal{N}(0,\Sigma_{\mathrm{e}, k})$ with band-diagonal matrix $\Sigma_{\mathrm{e}, k}$ consisting of elements:
\begin{align}\label{eq:cov-sigma-e}
    \E [\tilde e_{t, k} \tilde e_{t, k}^\top] =& G_{\mathrm{e},k} \mathrm{diag}_k(S)G_{\mathrm{e},k}^\top, \\
    \E [\tilde e_{t, k} \tilde e_{t+i,k}^\top] =& G_{\mathrm{e},k}  
    \begin{bmatrix}
        0 & 0\\ \mathrm{diag}_{k-i}(S) & 0 \notag 
    \end{bmatrix}
    G_{\mathrm{e},k}^\top, \ i\in \I{1,k-1},\notag \\
    \E [\tilde e_{t, k} \tilde e_{t+i,k}^\top] =& 0,\ i\in\I{k,T-t-k} \notag.
\end{align}
\changes{Given the Gaussian noise $\tilde{e}_{}[0:T-k],k$, the likelihood of the data satisfies}:
\begin{align}
    &\Pr  \left[\tilde e_{[0:T-k],k} {=} y_{[k:T]} - \Phi_{k, t} \theta_k\right]  \\
    \stackrel{\textcolor{blue}{\eqref{eq:e_defn}}}{\propto}&\exp \left[-\frac{1}{2}\|y_{[0:T-k]} - \Phi_{k, t} \theta_k\|^{\textcolor{blue}{2}}_{\Sigma_{\mathrm{e}, k}^{-1}}\right] \notag \\
    \stackrel{\eqref{eq:hat-thetak}}{=}&\exp \left[-\frac{1}{2}\|\theta_k - \hat \theta_k\|^{\textcolor{blue}{2}}_{\Sigma_{\theta, k}^{-1}}\right], \notag
\end{align}
Hence, $\hat \theta_k$ maximizes the likelihood function.
\end{proof}
Proposition~\ref{prop:id} presents a method for identifying multi-step predictors in their estimates. This approach utilizes the innovation form of the Kalman filter to construct a standard linear regression setup. By leveraging correlation information, we use \textit{generalized least-squares estimation} to obtain the MLE~\cite{Ljung1998}. Consequently, the proposed strategy inherits desired properties of the MLE framework. In particular, under mild assumptions it can be shown that the estimated parameter vector $\hat \theta_k$ is a \changes{asymptotically} consistent estimator of the true parameter vector $\theta_k$ and the errors $\changes{\theta_k-\hat \theta_k} \sim \mathcal{N}(0, \Sigma_{\theta, k})$ as $T \rightarrow \infty$~\cite{Ljung1998}. 

A key requirement for applying Proposition~\ref{prop:id} is the availability of $x_{t\mid t}$, $G_{\mathrm{e}, k}$, and $S$, computed using true state-space parameters. To bypass this, we propose estimating these entities via Kalman filter recursions with \changes{asymptotically} consistent estimates of the state-space matrices~\eqref{eq:state-space}, as described in Alg.~\ref{algo:solution}. An \changes{asymptotically} consistent estimate of the state-space model can be effectively obtained with existing solutions; e.g. Expectation-Maximization or Prediction Error Method\cite{GIBSON20051667, balim2024data, Ljung1998}. This approach simplifies the parameter identification process while ensuring that the error distribution $\theta_k - \hat{\theta}_k$ converges asymptotically to the same distribution as when using true system parameters~\cite{cox1979theoretical}. 
\changes{The following assumption introduces an approximation, assuming that the asymptotic Gaussian distribution of the MLE holds despite the utilization of a finite data length $T<\infty$.} 
\begin{assumption}\label{assum:param-dist}
    \changes{The multi-step predictor parameter estimates satisfy $\theta_k-\hat{\theta}_k \sim \mathcal{N}(0, \Sigma_{\theta, k})$, $\forall k \in \I{1,N}$.} 
Furthermore, the covariance matrices of the measurement and process noise are over-approximated, i.e., $\hat{R} \succeq R$ and $\hat G_\mathrm{w,k}\hat G_\mathrm{w,k}^\top \succeq G_\mathrm{w,k} G_\mathrm{w,k}^\top$. 
\end{assumption}
\changes{This approximation provides the basis for our controller design. We note that
similar approximations are common for data-driven control with stochastic noise~\cite{breschi2023data,yin2023stochastic}.} 
For the remainder of this paper, we assume that Assumption~\ref{assum:param-dist} holds and we denote $G_\mathrm{w,k} = \hat G_\mathrm{w,k}$, $R = \hat R$.

\begin{remark}[Related work]
    Identifying multi-step predictors is inherently challenging due to the increased number of parameters in the model and the noise correlations in subsequent multi-step transitions, as reflected in the noise covariance matrix~\eqref{eq:cov-sigma-e}. While this issue can be addressed with a simple least-square estimation~\cite{breschi2023data}, this necessitates a higher-order model, which reduces sample efficiency. In~\cite{yin2023stochastic, wang2022data}, a direct data-driven predictive control strategy is introduced, but it requires additional measurement of the disturbances/innovations. Alternatively, \cite{fiedler2023} present an asymptotically accurate method for identifying multi-step predictors. However, their approach assumes that the available data consists of independent samples rather than a trajectory consisting of correlated samples, which restricts its practical applicability. Use of a surrogate state-space model estimate, as proposed in~\cite{koehler2022state}, can address this issue by allowing disjoint estimation of multi-step predictors in systems with noisy state measurements. We extend this strategy to systems with partial measurements through the innovation form of the Kalman filter. Our identification strategy, therefore, imposes milder practical assumptions compared to existing literature, as it can effectively utilize data from a single input-output trajectory.
\end{remark}

\section{Stochastic Predictive Control with data-driven Multi-Step Predictors}\label{sec:control}
In the following, we reformulate Problem~\eqref{eq:soc} using the uncertain multi-step predictors derived in Section~\ref{sec:id}. To jointly address the stochastic uncertainty in the model parameters $\theta_k$ and the future disturbances $w_{[0,N]}$, we employ the concept of \textit{robustness in probability}~\cite[Lemma V.3]{wabersich2021probabilistic}. Specifically, we construct sets $\Theta_{k,j}$ such that $\Pr[\theta_k \in \Theta_{k,j}] \geq \delta$, where $\delta \in (p, 1)$ is a user-chosen\footnote{\changes{A natural choice is $\delta=\tilde{p}=\sqrt{p}$.}} probability. Then, we enforce the chance constraints~\eqref{eq:chance-constr} robustly for all parameters $\theta \in \Theta_{k,j}$ with an increased probability of $\tilde p = p/\delta$. The following proposition recaps a simple approach to address this problem by using the standard confidence ellipsoids $\Theta_{k,j}$:
\begin{proposition}[{Adapted from~\cite[Lemma~3]{koehler2022state}}]\label{prop:ellipsoidal}
    Let Assumption~\ref{assum:param-dist} hold and suppose
    \begin{align}\label{eq:ct-ell}
         &h_j^\top[\hat G_{0, k},\ \hat G_{\mathrm{u}, k}]\begin{bmatrix} \bar x_0 \\ u_{[0,k-1]} \end{bmatrix}+ \changes{d_{\delta, k}}\left\|{\Sigma_{\mathrm{\theta}, k}^{1/2}\left(\begin{bmatrix} \bar x_0 \\ u_{[0,k-1]}\end{bmatrix}\otimes h_j\right)}\right\|  \notag \\ 
        &\leq  1- c_{\tilde p}\sqrt{d_{k,j} + f_{k,j}}, \; \; j \in \mathbb{I}_{[1, r]},\ k \in \I{1,N},
    \end{align}
    with $ \changes{d_{\delta, k}} =\sqrt{\chi^2_{n_{\theta,k}}(\delta)}$, $  d_{k,j} = h_j^\top (G_{\mathrm{w}, k}G_{\mathrm{w}, k}^\top + R)h_j$, 
    \begin{align}\label{eq:constants}
        f_{k, j} &= 
        \max_{\theta_k:\|\theta_k - \hat \theta_k\|_{\Sigma_{\theta,k}}\leq \changes{d_{\delta, k}}}\theta_k^\top M_{k,j} \theta_k,\\
\label{eq:M}
        M_{k,j} &= \left(\begin{bmatrix}
                \Sigma_{\mathrm{x}, 0} & 0 \\ 0 & 0
            \end{bmatrix}\otimes h_j h_j^\top \right).
    \end{align}
    Then, applying $u_{[0, N-1]}$ satisfies the chance constraints~\eqref{eq:chance-constr}.
\end{proposition} 
Proposition~\ref{prop:ellipsoidal} first establishes  confidence ellipsoids $\Theta_{k,j}$ using the provided Gaussian distribution and then derives tight upper bounds for the predicted output. However, when the parameter dimension $n_\theta$ is large, the resulting confidence ellipsoids become \changes{overly conservative}. Instead, we constructs tight sets $\Theta_{k,j}$ separately that are independent of the number of parameters. In particular, we improve the constraint tightening~\eqref{eq:ct-ell} with the following optimization problem:
\begin{subequations}
\label{eq:soc-proposed}
        \begin{align}
         \min_{u_{[0, N-1]} \in \mathbb{U}^N}& \sum_{k=0}^{N-1}\| \hat y_{k+1}\|_{Q_\mathrm{c}}^2 + \|u_k\|_{R_\mathrm{c}}^2 \\
        \text{s.t. } &  h_j^\top \hat y_k+c_{\epsilon} \left\|{\Sigma_{\mathrm{\theta}, k}^{1/2}\left(\begin{bmatrix} \bar x_0 \\
        u_{[0,k-1]}\end{bmatrix}\otimes h_j\right)}\right\|   \label{eq:tightened_constr}\\ 
        &\leq  
1- c_{\tilde p}\sqrt{f_{\epsilon, k, j} + d_{k,j}},\; j \in \mathbb{I}_{[1, r]},\; k \in \I{1,N}, \notag \\
    &\hat y_k= \hat G_{\mathrm{0},k} \bar x_0 +  \hat G_{\mathrm{u},k} u_{[0,k-1]},\ k\in\I{1, N}, \label{eq:hatyk}
        \end{align}
    \end{subequations}
    with $c_\epsilon = \sqrt{\chi_1^2(2\epsilon-1)}$ and $f_{\epsilon, k, j}$ according to the percentile function of generalized chi-squared distribution~\cite{genchi-davies}:
    \begin{align}\label{eq:genchi}
        f_{\epsilon, k, j} &= \min_f f \\
        \text{s.t.} &\ \Pr[\theta_{k}^\top  M_{k,j} \theta_{k} \leq f,~\theta_k\sim\mathcal{N}(\hat{\theta}_k,\Sigma_{\theta,k})] \geq 1+\delta-\epsilon. \notag
    \end{align}
     \changes{The optimization problem~\eqref{eq:soc-proposed} is a convex second-order cone program, where the linear constraints of the nominal problem~\eqref{eq:soc-msp} are replaced by second-order cone constraints that ensure chance constraint satisfaction under parametric uncertainty.}
    The following theorem establishes satisfaction of the considered chance constraints~\eqref{eq:chance-constr}:
\begin{theorem}\label{thm:main}
    Consider Assumption~\ref{assum:param-dist} holds, and any input sequence $u_{[0, N-1]}$ satisfying the constraints in Problem~\eqref{eq:soc-proposed}. Then, applying $u_{[0, N-1]}$ ensures satisfaction of the chance constraints~\eqref{eq:chance-constr}.
\end{theorem}
\begin{proof}
Using properties of Kronecker product, we can establish the following relation
\begin{align}
     h_j^\top \bar y_k \stackrel{\eqref{eq:baryk}}{=}\theta_{k}^\top\left(\begin{bmatrix} \bar x_0 \\ u_{[0,k-1]}\end{bmatrix}\otimes h_j\right). 
\end{align}
Thus, Inequalities~\eqref{eq:tightened_constr},\eqref{eq:hatyk} and $\theta_k\sim\mathcal{N}(\hat{\theta}_k,\Sigma_{\theta,k})$ imply:
\begin{align}\label{eq:pf1}
    \Pr \bigg [h_j^\top\bar y_k  & \leq 1 -  c_{\tilde p} \sqrt{f_{\epsilon, k, j} + d_{k,j}} \bigg] \geq \epsilon.
\end{align}
Next, we derive the covariance propagation for the initial state using Kronecker product properties:
\begin{align}
\label{eq:proof_M}
    &\|\Sigma_{\mathrm{x}, 0}^{1/2}G_{0, k}^\top h_j\|^2  \notag
    =\left \|\begin{bmatrix}
            \Sigma_{\mathrm{x}, 0}^{1/2} & 0 \\ 0 & 0
        \end{bmatrix}\begin{bmatrix}
            G_{0, k}^\top \\ G_{\mathrm{u}, k}^\top
        \end{bmatrix} h_j \right \|^2 \notag\\
    =& \left \|\begin{bmatrix}
             \Sigma_{\mathrm{x}, 0}^{1/2} & 0 \\ 0 & 0
        \end{bmatrix} (I\otimes h_j)^\top \theta_k \right \|^2 
    \stackrel{\eqref{eq:M}}{=}\theta_k^\top M_{k,j} \theta_k.
\end{align}
Consequently, using the covariance dynamics~\eqref{eq:covar-dyn}:
\begin{align}
    \|h_j\|_{\Sigma_{\mathrm{y},k}}^2 &= h_j^\top \Sigma_{\mathrm{y},k} h_j \\
    &= h_j^\top (G_{0,k} \Sigma_{\mathrm{x},0}G_{0,k}^\top + G_{\mathrm{w},k}G_{\mathrm{w},k}^\top + R) h_j \notag \\
    &= \theta_k^\top M_{k,j} \theta_k + d_{k,j} \notag
\end{align}
Using the definition of $f_{\epsilon, k,j}$ from Eq.~\eqref{eq:genchi}:
\begin{equation}\label{eq:barc}
    \Pr[ \|h_j\|_{\Sigma_{\mathrm{y},k}}\leq \sqrt{f_{\epsilon, k, j} + d_{k,j}}] = 1+\delta - \epsilon.
\end{equation}
Last, we combine~\eqref{eq:pf1} and~\eqref{eq:barc} with Boole's inequality to show that $\Pr[\theta_k \in \Theta_{k,j}] \geq \delta$ where:
    \begin{equation}\label{eq:robust-set}
        \Theta_{k,j} =\left \{\theta_k \ \middle | \  h_j^\top\bar y_k  \leq 1 -  c_{\tilde p} \|h_j\|_{\Sigma_{\mathrm{y}, k}}\right \}
    \end{equation}
    Finally, $\theta_{k}\in \Theta_{k,j}$ implies the satisfaction of constraints~\eqref{eq:chance-constr} with probability $\tilde p$ since $y_k\sim\mathcal{N}(\bar{y}_k,\Sigma_{\mathrm{y},k})$. Consequently,
    \begin{equation*}
        \Pr [ h_j^\top y_k  \leq 1  \mid  \theta_k \in \Theta_{k,j}] \Pr[\theta_k \in \Theta_{k,j}] \geq \tilde p  \delta = p. \qedhere
    \end{equation*}
\end{proof}

\changes{The tightened constraints~\eqref{eq:tightened_constr} in the proposed control problem provide a sufficient condition for chance constraint satisfaction. The term $d_{\delta, k}$ in~\eqref{eq:ct-ell} grows with the parameter dimension $n_{\theta,k}$, increasing conservatism with a larger horizon. In contrast, using a parameter set represented by a single half-space constraint, this term is replaced by $c_\epsilon$ in Problem~\eqref{eq:soc-proposed}, which is independent of parameter dimension $n_{\theta,k}$ and horizon $k$, reducing conservatism.}. The only conservatism introduced arises from using union bound for combining uncertainty due to initial state's mean and covariance. Finally, the proposed tightening strategy can be easily extended to handle two-sided constraints by replacing the constants $c_{\epsilon}$ and $c_{\tilde p}$ with their two-sided counterparts~\cite{kouvaritakis2016model}.

\section{Numerical Example}\label{sec:example}
In the following, we illustrate the proposed framework using a mass-spring-damper chain. Computations were conducted in Python on a server instance with 16 GB RAM and 8-core allocation from an AMD EPYC 9654 96-core processor. The optimization problems are solved via ECOS~\cite{domahidi2013ecos}. The numerical example's code is accessible online at: \url{https://github.com/haldunbalim/MultistepPredictors}.

\subsubsection*{Problem setup}
We study a chain of $3$ mass-spring-damper system with the control input affecting the last mass, resulting in a configuration where $n_\mathrm{x} = 6$ and $n_\mathrm{u} = 1$. The system parameters are chosen as: mass $1 \, \text{kg}$, spring constant $10 \, \text{N/m}$, and damping $2 \, \text{kg/s}$. The disturbance covariance is $EE^\top= \mathrm{diag}(0_3,\ 10^{-3}I_3)$. Only noisy position measurements are available, $n_\mathrm{y} = 3$, with measurement noise covariance $R = 10^{-3}I$. System matrices are derived via exact discretization with a $0.5 \, \mathrm{s}$ step and converted to observer canonical form. The user chosen parameters for the control problem~\eqref{eq:soc-proposed} are set as: $N=20$, $p=0.9$, $\delta=0.95$, $\epsilon=0.975$.

\subsubsection*{System Identification}
To estimate system models, we obtained a trajectory of length $T=10^3$ by exciting the system with random normal inputs $u_t \sim \mathcal{N}(0,4)$. The state-space model is estimated using the constrained EM algorithm from~\cite{balim2024data}, with matrices $E$ and $R$ known up to a scaling constant. The computation time for state-space model identification is $11.5$\,s, while for all multi-step predictors, it is additionally $1.09$\,s.

\subsubsection*{Probabilistic Reachable Set}
First, we assess the conservatism of the proposed constraint tightening strategy~\eqref{eq:tightened_constr} by computing the probabilistic reachable set for $u \equiv 5$. The initial state covariance is set to the steady-state covariance derived from the actual system parameters. In this section, we compare the following methods:

\begin{enumerate}\label{enum:methods}
    \item \textit{Sequential}: Sequential propagation using state-space model and homothetic tubes~\cite{balim2024data};
    \item \textit{Ellipsoidal}: Multi-step predictor with confidence ellipsoids  (Proposition~\ref{prop:ellipsoidal});
    \item \textit{Proposed}: Multi-step predictor using the derived uncertainty characterization directly (Theorem~\ref{thm:main});
    \item \textit{Sampling}: Estimate the probabilistic reachable set using $10^5$ samples $\theta_k\sim\mathcal{N}(\hat{\theta}_k,\Sigma_{\theta,k})$ for each $k\in\mathbb{I}_{[1,N]}$.
\end{enumerate}

\begin{figure}
\begin{center}
\includegraphics[width=0.9\columnwidth]{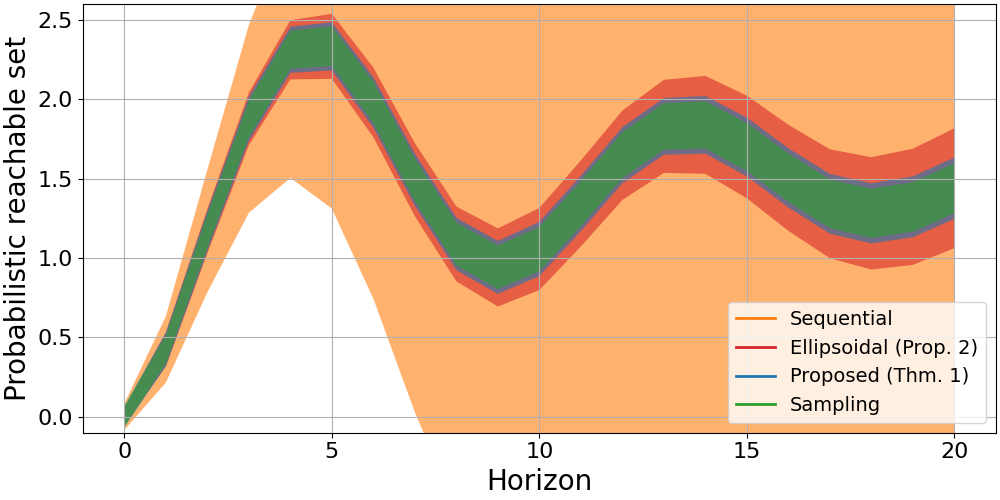} 
\caption{Comparison of the probabilistic reachable sets for the last mass's position measurements for $N=20$.}
\label{fig:fwd_reach}                                 
\end{center}                               
\end{figure}

In Figure~\ref{fig:fwd_reach}, we show two-sided probabilistic reachable sets for the last mass's position measurements with probability level $p=0.9$. 
The sequential propagation results in overly conservative estimates while the multi-step predictors closely approximate the ground truth. 
Furthermore, the proposed strategy is less conservative than the ellipsoidal set-based approach, since we directly use distribution of parameters. Finally, the small difference between our approach and the sampling-based estimate of the probabilistic reachable set 
highlights the \changes{reduced} conservatism of our approach.

\subsubsection*{Predictive Control}
In the following, we highlight the ability of the proposed framework to enforce the chance constraints~\eqref{eq:chance-constr}. We simulate the system and estimate state-space and multi-step models for $10^3$ times. The initial state is set with a mean $\bar x_{0} \equiv -0.2$ and a covariance matching the steady-state derived from the true system parameters. Input constraints are set as $|u_{t}| \leq 2.5$ and chance constraints on position measurements are set as $\Pr [y_{t,i} \leq 0.05] \geq p=0.9$. We consider the cost matrices $Q_{\mathrm{c}} = I$ and $R_{\mathrm{c}} = 10^{-1}I$. Table~\ref{tab:feas} shows the number of instances where the considered methods find feasible solutions \changes{and offline design time including state-space identification}. As shown, the multi-step methods significantly outperform the sequential method in this example, in accordance with Fig.~\ref{fig:fwd_reach}. Moreover, the proposed constraint tightening strategy yields feasible solutions in significantly more cases compared to the strategy using confidence ellipsoids.

\begin{table}[]
\caption{Number of instances where the control problem is feasible and offline computation time required for each method.}
\centering
\begin{tabular}{lcc}
\hline
Method      & \makecell{Comp.  Time(s)} & Feasible (\%) \\
\hline
Sequential  & 44.9  & 0.0  \\ 
Ellipsoidal & 25.9  & 58.9   \\
Proposed    & 33.5  & 99.2   \\
\hline
\end{tabular}
\label{tab:feas}
\end{table}

Finally, we compare the computational time, and constraint satisfaction of the proposed scheme against the stochastic MPC~\eqref{eq:soc-msp} with true system parameters. Out of $10^3$ trials, we neglect $8$ instances where problem~\eqref{eq:soc-proposed} was infeasible. We compute $f_{\epsilon, k, j}$ in Eq.~\ref{eq:genchi} using Davies' method~\cite{genchi-davies} and approximate it using sampling if numerical issues arise. \changes{
The increased computational time of the proposed approach~\eqref{eq:soc-proposed} is due to the second order cone constraints that address the parametric uncertainty}. Results for this experiment are presented in Table~\ref{tab:stats}. As shown, the proposed strategy successfully enforces the desired chance constraints. \changes{The proposed approach achieves constraint violations below $10\%$, highlighting the guarantees of Theorem~\ref{thm:main} and its conservatism relative to the idealized solution.}

\begin{table}[]
\caption{Comparison of maximum constraint violation probability over all time steps, and average computation time.}
\begin{center}
\begin{tabular}{lcc}
\hline
Method   & \makecell{Time for \\ Control Prob. (ms)} & \makecell{Constr. \\Violation (\%)} \\
\hline
Exact model (idealized)~\eqref{eq:soc-msp}        &       5.1         &      10.0                \\
Proposed (\changes{robust})~\eqref{eq:soc-proposed}   &        40.4        &      3.7                \\
\hline
\end{tabular}
\label{tab:stats}
\end{center}
\end{table}

Overall, the numerical example shows that the proposed method significantly reduces conservatism compared to state-of-the-art solutions while reliably satisfying the probabilistic constraints despite the lack of a precise model knowledge.

\section{Conclusion}\label{sec:conclusion}
We presented a data-driven framework for addressing the stochastic optimal control problem using multi-step predictors. We first propose an identification and uncertainty quantification scheme for multi-step predictors utilizing an intermediate state-space identification step. The identified multi-step predictors are used to directly propagate uncertainty in the parameter estimates, allowing for \changes{simple constraint tightening strategies that ensure chance constraint satisfaction}. Finally, we provided a numerical example, highlighting the ability to guarantee satisfaction of probabilistic constraints while significantly reducing conservatism compared to state-of-the-art approaches. Future work will focus on developing a framework that addresses the receding horizon case with closed-loop guarantees, by incorporating ideas from~\cite{balim2024data, saccani2023homothetic}. \changes{Extending the proposed techniques to nonlinear and unstable systems is an interesting avenue for research.}

\bibliographystyle{IEEEtran}
\bibliography{final}
\end{document}